\documentclass[%
 reprint,
 amsmath,amssymb,
 aps,
 pre,
 longbibliography
]{revtex4-1}

\usepackage{CJK}
\usepackage{graphicx}
\usepackage{dcolumn}
\usepackage{bm}
\usepackage{color}
\usepackage{amsmath}

\begin{document}



\title{Nucleation of Spatio-Temporal Structures From Defect Turbulence in the Two-dimensional Complex Ginzburg--Landau Equation}

\author{Weigang Liu} \email{qfsdy@vt.edu}
\author{Uwe C. T\"auber} \email{tauber@vt.edu}
\affiliation{Department of Physics (MC 0435) and Center for Soft Matter and Biological Physics, Robeson Hall, 850 West Campus Drive, Virginia Tech, Blacksburg, VA 24061, USA}

\date{\today} 

\begin{abstract}
We numerically investigate nucleation processes in the transient dynamics of the two-dimensional complex Ginzburg--Landau equation towards its ``frozen'' state with quasi-stationary spiral structures. 
We study the transition kinetics from either the defect turbulence regime or random initial configurations to the frozen state with a well-defined low density of quasi-stationary topological defects. 
Nucleation events of spiral structures are monitored using the characteristic length between the emerging shock fronts. 
We study two distinct situations, namely either when the system is quenched far away from the transition limit or near it. 
In the former deeply quenched case, the average nucleation time for different system sizes is measured over many independent realizations. 
We employ an extrapolation method as well as a phenomenological formula to account for and eliminate finite-size effects.
The non-zero (dimensionless) barrier for the nucleation of single spiral droplets in the extrapolated infinite system size limit suggests that the transition to the frozen state is discontinuous. 
We also investigate the nucleation of spirals for systems that are quenched close to but beyond the crossover limit, and of target waves which emerge if a specific spatial inhomogeneity is introduced. 
In either of these cases, we observe long, ``fat'' tails in the distribution of nucleation times, which also supports a discontinuous transition scenario.
 
\end{abstract}

\maketitle


\section{Introduction} \label{sec:intro}

Spontaneous spatial structure formation, as well as its inverse process, the dynamical destruction of patterns through spatio-temporal chaos, are of considerable interest, especially in far-from-equilibrium systems. 
These phenomena are encountered in condensed matter physics, chemistry, and biology. 
Studies of this topic have benefited from the recent development of careful experiments, as well as analytic and numerical tools. 
Non-equilibrium spatial patterns often emerge due to linear system instabilities that are induced by varying some control parameter(s) beyond certain thresholds \cite{Cross93, Cross09}. 
The term ``ideal pattern'' is defined to represent the spatially periodic structure of an infinitely extended system. 
Defects, which can generally be viewed as any departure from this ideal pattern, constitute important ``real pattern'' effects. 
Their structures may reflect the topological characteristics of the ideal patterns in a (quasi-)stationary dynamical state. 
For example, the (quasi-)stationary kinetics in non-equilibrium steady states as well as the relaxation towards such (quasi-)stationary regimes are often governed by the properties of topological defects in these non-linear dynamical systems.
Indeed, the onset of spatio-temporal chaos in excitable or oscillatory media can trigger the appearance of topological defects, which are point-like objects in two dimensions, but can be extended in higher dimensions. 
A striking example are the famous chemical oscillations in the Belousov--Zhabotinsky (BZ) reactions \cite{Zaikin70, Winfree72, Winfree73, Muller87, Skinner91}, where the defect points emit spiral chemical waves. 
Such striking wave patterns are also observed in spatially extended stochastic population dynamics, e.g., the stochastic May--Leonard model with three cyclically competing species \cite{Reichenbach06, Reichenbach07, He11, Serrao17}. 
Other types of topological defects are encountered in driven systems maintaining non-linear traveling waves such as Rayleigh--B\'enard convection in planar nematic liquid crystals \cite{Daviaud92, Feng92}, or electroconvecting nematic liquid crystals \cite{Rehberg89}. 
Here the defects constitute dislocations in the roll pattern of the traveling waves \cite{Coullet88}.

A simple, generic description of many pattern-forming systems is afforded through the complex Ginzburg--Landau equation (CGL). 
Indeed, the CGL is considered to at least represent the ``kernel'' of many amplitude equation models \cite{Chate96} that have been employed to characterize spontaneous spatial pattern formation \cite{Coullet88, Komineas00}. 
A large number of numerical studies of the CGL in two dimensions has been performed over the past three decades, and uncovered a wide variety of intriguing dynamical behavior upon varying certain control parameters. 
In this work, we mostly consider the transition from the ``defect-mediated turbulence'' state with strong spatio-temporal fluctuations to the so-called ``frozen'' state displaying beautiful quasi-stationary spiral wave patterns that resemble structures observed in two-dimensional oscillating media (e.g., the BZ reactions and the stochastic May--Leonard model). 
Only a few previous studies have addressed the transition dynamics between these two states, namely the crossover from the strongly fluctuating into a dynamically frozen state \cite{Huber92}. 
This process is thought to be important for controlling spatio-temporal chaos in spatially extended systems \cite{Aranson94, Zhang02}. 
The inverse transition has been also observed in a real BZ reaction with an open reactor \cite{Zhang03}. 
The transition between turbulent and frozen spiral states can be altered by adding a chiral symmetry breaking term into the right-hand-side of the CGL \cite{Nam98}, which also modifies the specific form of the dynamical equation and hence the phase diagram. 
To obtain a better understanding of the accompanying transient processes, we aim for a quantitative characterization of the transition dynamics, where one switches the control parameters beyond a transition limit in parameter space. 
The defect-mediated turbulence state is thus rendered meta-stable, and well-established spiral structures emerge. 
As pointed out in Ref.~\cite{Huber92}, when turbulence becomes transient, the onset of the formation of spiral wave structures can be viewed as a nucleation process. 
The authors of Ref.~\cite{Huber92} primarily investigated situations where the control parameters were chosen near the crossover regime, and hence the defect-mediated turbulence state becomes just unstable. 
In the present study, we conversely also consider sudden parameter switches which represent deep quenches into the stable frozen state regime. 
Based on our numerical results and subsequent finite-size extrapolation analysis, we obtain evidence for a non-vanishing nucleation barrier for the formation of spiral structures.
Consequently we conclude that the transition from the defect turbulence to the quasi-stationary frozen state is of a discontinuous nature.

In addition to spiral wave structures, reaction-diffusion systems may also exhibit ``target wave patterns'' in the presence of spatial inhomogeneities. 
For example, the BZ reaction supports target waves if a dust grain or other impurity is inserted \cite{Zaikin70, Winfree73}. 
According to Refs.~\cite{Hagan81, Kopell81a}, a pacemaker is needed to stabilize the resulting wave pattern. 
However, target patterns can also emerge in a homogeneous system if it is subjected to perturbations with oscillating concentrations \cite{Kopell81a, Kopell81b, Hagan81, Bugrim96}. 
The singularity point in the defect center will assume this trigger role for stable spiral structures, whence spontaneous creation and annihilation of defect pairs may occur.
Those pacemakers or nucleation centers ensure that the internal regions of those heterogeneous nuclei will maintain an effectively higher oscillation frequency than the outside bulk of the oscillatory medium, which is the reason why a local inhomogeneity is needed to generate target waves. 
However, in contrast to rotating spirals, target waves take a concentric circular shape, and propagate radially outward from their source. 
Previous work indicated that target waves are characterized by a vanishing topological charge \cite{Hendrey00a}, in distinction to the positively or negatively charged (oriented) spirals. 
There are also intriguing studies of spatiotemporal chaos control through target waves \cite{Jiang04}. 
Thus, in this work we also study target wave nucleation and compare this scenario with the aforementioned spiral droplet nucleation.

This paper is structured as follows: 
A general review of the CGL model is presented in section~\ref{sec:model}; section~\ref{sec:scheme} describes our numerical scheme as well as the methodology we devised to characterize the spatial scale of the CGL system, especially when it contains ``droplet'' nucleating structures, and to determine the nucleation time distribution. 
In section~\ref{sec:spiral_nuc}, we implement our algorithm and numerical analysis to study spontaneous spiral droplet nucleation events in two-dimensional CGL systems subject to two distinct quench protocols: namely either from fully randomized initial conditions, or from initial configurations that correspond to stationary states in the defect turbulence region. 
Target wave nucleation with an artificially prepared pacemaker are investigated and analyzed in section~\ref{sec:target_nuc}. 
Finally, we summarize our results in the concluding section~\ref{sec:conclusions}.

\section{Model Description} \label{sec:model}
A general description of non-linear driven-dissipative systems is afforded by the complex Ginzburg--Landau equation in terms of a complex field $A(\bm{x},t)$ \cite{Newell69, Kuramoto84, Newell93}, namely the complex amplitude  related to a characteristic quantity $C(\bm{x},t)$ describing the system. 
Therefore, $C(\bm{x},t)$ can be decomposed as:
\begin{equation}
    C(\bm{x},t) = C_0 + A(\bm{x},t) e^{i\omega_0 t}\bm{U_l} + c.c. + h.o.t. ,
    \label{chemode}
\end{equation}
where $C_0$ is a constant, $c.c.$ means complex conjugation, $h.o.t.$ indicates higher-order terms, $\bm{U_l}$ denotes the eigenvector for the linear instability, and $\omega_0$ the corresponding eigenvalue. 
Furthermore, compared with the frequency scale $\omega_0$, the complex field $A(\bm{x},t)$ is assumed to be slowly varying. 
$A(\bm{x},t)$ is often called order parameter, and obeys the CGL which can be written in a rescaled form:
\begin{eqnarray}
    \partial_t A(\bm{x},t) &=& A(\bm{x},t) + (1+ib) \nabla^2 A(\bm{x},t) \nonumber \\
    &&- (1+ic)| A(\bm{x},t)|^2 A(\bm{x},t) ,
    \label{cglec1}
\end{eqnarray}
with real constants $b$ and $c$ that characterize the linear and non-linear dispersion, respectively. 
Here, the parameters that describe the deviation from the transition threshold, the diffusivity, and the non-linear saturation have been rescaled to $1$. 
Note that the frequency shift (the imaginary part of the coefficient of the linear term) is eliminated by gauge symmetry. 
Eq.~(\ref{cglec1}) is invariant under the transformation $(A,b,c) \to (A^*,-b,-c)$ \cite{Aranson02}. 
Therefore, we only need to consider half of the parameter space in the $b$-$c$ plane and can directly predict same behavior of the system from this mapping; for example, in the two-dimensional $(b,c)$ parameter space, the first and third quadrants are equivalent due to that invariance transformation. 
The first and third quadrants are labeled ``defocusing quadrants'', since the associated spiral waves rotate along the same direction as the equi-phase lines (or spiral arms). 
The second and fourth quadrants are correspondingly called ``focusing quadrants'', and yield spirals that rotate inversely with respect to the defocusing quadrants. 
We shall restrict ourselves to the focusing case in this paper.

The CGL (\ref{cglec1}) describes spatially extended systems whose homogeneous state is oscillatory around the threshold of a super-critical Hopf bifurcation, e.g., for which the stable stationary dynamical state becomes a global limit cycle. 
The isotropic non-linear partial differential equation~(\ref{cglec1}) reduces to the ``real'' Ginzburg-Landau Equation (GL) for $b = c =0$. 
It may also be viewed as a dissipative extension of the conservative non-linear Schr\"odinger equation \cite{Aranson02}, which formally follows in the limits $b, c \to \infty$ in Eq.~(\ref{cglec1}).

There exists a simple homogeneous solution of Eq.~(\ref{cglec1}), namely $A = \exp(-ict + \phi)$ with frequency $\omega = c$ and an arbitrary constant phase $\phi$. 
This spatially uniform periodic solution becomes unstable when $1 + bc < 0$, known as the Benjamin--Feir limit. 
Long-wavelength modes with wave numbers below a critical threshold proportional to $\sqrt{|1 + bc|}$ will then be exponentially enhanced. 
A more general solution form is a family of traveling plane wave solutions,
\begin{equation}
    A(\bm{x},t) = \sqrt{1-Q^2} \, e^{i (\bm{Q \cdot x} - \omega t)},
    \label{plan_solu}
\end{equation}
with frequency $\omega = c + (b-c) Q^2$, restricted to $Q^2 < 1$. 
The homogeneous oscillating solution is restored in the limit $Q \to 0$.
In order to test the stability of this solution family, the complex growth rates $\lambda$ of the perturbed modulational modes are determined perturbatively, and the associated instabilities can be inferred from their real parts \cite{Aranson92a, Weber92, Aranson02}.
Restricting the analysis to the most dangerous longitudinal perturbation and performing a long-wavelength expansion (with $b \ne c$ and $Q \ne 0$), one arrives at the Eckhaus criterion for the plane wave solution (\ref{plan_solu}) that can be tested against convective instability for non-zero group velocities \cite{Kramer85, Tuckerman90, Janiaud92}, for which the initial localized perturbation will be amplified. 
However, at fixed position, it can in fact not be amplified due to the drift \cite{Landau59}. 
The absolute instability limit is obtained by considering the evolution of a localized perturbation in the linear regime, given by
\begin{equation}
    S(\bm{x},t)=\int \frac{d^dk}{(2\pi)^d} \hat{S}_0(\bm{k}) e^{i\bm{k\cdot x} + \lambda(\bm{k}) t},
    \label{ab_instable}
\end{equation}
where $\hat{S}_0(\bm{k})$ denotes the Fourier transform of the initial perturbation $S_0(\bm{x})$, and $\lambda(\bm{k})$ represents the growth rate of specific modulational modes with wave vector $\bm{k}$ \cite{Stuart80}. 
In the asymptotic time limit $t\to \infty$, the integral will be dominated by the largest saddle point of the growth rate $\lambda(\bm{k})$. 
The criterion of absolute instability finally is given by
\begin{equation}
    {\rm Re}[\lambda(\bm{k}_0)] > 0 , \ \partial_{\bm{k}}\lambda(\bm{k})|_{\bm{k}_0} = 0 .
    \label{ab_instable_cri}
\end{equation}
It suggests that the Eckhaus instability is more restrictive than the absolute stability limit when $Q \ne 0$ \cite{Weber92}. 
``Spatio-temporal'' chaos, which has received continuous interest over the past decades, is obtained upon moving beyond those limits into the unstable regime in the $(b-c)$ parameter plane.

In order to specifically describe the dynamics in the phase-unstable regime $1 + bc < 0$ and near the bifurcation threshold, one may just consider the most unstable modes \cite{Coullet89b}; namely, only the phase term constitutes a relevant order parameter when its gradients remain sufficiently small. 
The emerging ``phase turbulence'' regime is characterized by relatively weaker spatio-temporal chaos without the presence of topological defects (see below). 
Manneville and Chat\'e characterized the statistical properties of this state through evaluating the parameters of an effective Kardar--Parisi--Zhang equation \cite{Manneville96}. 
However, this reduced description breaks down in the limit $1 + bc \ll -1$ \cite{Coullet89a}. 

Here, a turbulent state with strong coupling between the amplitude and phase modes emerges, named ``amplitude turbulence'', and governed by exponential decays of the correlation function in both time and distance; therefore, no long-range temporal or spatial order persists. 
The existence of hysteretic behavior observed in Refs.~\cite{Coullet89a, Sakaguchi90, Shraiman92} as well as the Lyapunov exponent measured through different methods \cite{Bohr90} suggest that the transition from a homogeneous periodic solution to this amplitude turbulence regime is discontinuous. 
The appearance of the turbulent state is accompanied by the presence of topological defects, which will take the form of points in two spatial dimensions, and defect lines in three dimensions, and are located at the points where $|A(\bm{x},t)| = 0$, or the phase gradient of $A(\bm{x},t)$ diverges \cite{Bohr90}. 
They can be characterized by the quantized circulation
\begin{equation}
    \oint_C d\theta = 2\pi n , \ n = \pm1, \pm2, \ldots ,
    \label{defect0}
\end{equation}
with $\theta = \arg A$, and where $C$ denotes an arbitrary closed path which encloses the core of a defect, and the integer n represents a ``topological charge''. 
Previous studies suggest that multiply charged defects are unstable and will split into a set of single-charged defects \cite{Hagan82a}. 
The mechanism of defect creation was described in earlier work as well \cite{Coullet89a}. 
In the phase-unstable regime, the phase turbulence quickly saturates, and phase gradients increase rapidly. 
This will eventually lead to a pinching of the equi-phase, followed by a shock-like event. 
This sequence hence creates a pair of topological defects with opposite charge \cite{Gaponov87}. 
Since in the ultimate well-established turbulent state, the system will tend to have a uniform distribution of topological charge, these creation events are likely to happen ``far from'' any existing defects, implying a constant defect generation rate. 
On the other hand, defect pair annihilation processes can be understood on a mean-field level assuming random defect motion. 
The annihilation rate will then be roughly proportional to the square of the defect number inside the system \cite{Gil90, Huber92}. 
Finally, the ``defect-mediated turbulence'' regime describes a stationary configuration with defects continuously appearing, moving, mixing, and annihilating \cite{Lega91, Coullet89a, Gil90, Coullet89b, Wang04}; it is consequently characterized by the balance of spontaneous creation and annihilation of topological defects.

In the special case $b = c$, the dynamics of the system described by Eq.~(\ref{cglec1}) will resemble the critical dynamics of ``model E'' \cite{Hohenberg77}  for a non-conserved two-component critical order parameter field, e.g., in a planar ferromagnet or superfluid in equilibrium \cite{Tauber14}, and restores the ``real'' GL. 
In two dimensions, it yields the dynamics of the Berezinskii--Kosterlitz--Thouless transition observable, e.g., in superfluid helium films \cite{Bishop80}. 
The topological defects will take the form of vortices as in the planar XY model; if $b = c \ne 0$, the vortices will rotate.
When $b \ne c$, topological defects may emit spiral waves whose arms (the equi-phase lines) behave as those of an Archimedean spiral \cite{Skinner91}; they can either propagate inward or outward. 
Those spiral wave structures are thought to be very important features in biological systems \cite{Davidenko91}. 
Stable spiral waves can be formed beyond a certain limit in the $b$-$c$ plane where the turbulent state becomes metastable. 
The existence of localized amplitude modes, i.e., stable topological defects, will allow nucleation to happen, and thereby eventually generating stable spiral structures in the system.

Aside from directly varying the control parameters $(b,c)$, one may also change the two-dimensional CGL system behavior by introducing suitable spatially localized inhomogeneities, which will modify Eq.~(\ref{cglec1}) by adding a local perturbation on its right-hand-side. 
This additional heterogeneity can facilitate the formation of so-called ``target wave'' patterns. 
Specifically, if the local oscillation frequency in the perturbed spatial patch is set to be lower than in the surrounding bulk, and if the homogeneous solution of Eq.~(\ref{cglec1}) has a stable limit cycle, there should be a unique solution independent of the initial condition as $t \to \infty$, which is comprised of radially outward propagating waves that originate from the localized inhomogeneity \cite{Hendrey00b}.
Asymptotically, those spreading fronts should behave just like plane waves (\ref{plan_solu}). 
According to Hagan's theoretical study \cite{Hagan81}, there actually exists a unique and stable heterogeneous target wave solution of the modified CGL (\ref{cglec1}) with an additional inhomogeneity. 
Its most common form is introduced by setting the control parameters $(b,c)$ within a small region different from its surrounding environment. 
Furthermore, two different types of target wave solutions, distinguished by an additional superimposed temporal modulation at another frequency in one of them, can be obtained if a more complicated spatial inhomogeneity is introduced in the system \cite{Hendrey00b}. 
Target wave solutions can also be generated by boundary effects \cite{Eguiluz99}.

When wave fronts emitted by different sources, e.g., topological defects or even heterogeneous nuclei, collide, their interaction causes the appearance of a ``shock'', i.e., in two dimensions a thin linear structure in the modulus field. 
These shocks are considered to be very strong perturbations marked by a sudden increase of $|A|$.
As they can absorb incoming perturbations, interactions between waves originating from different sources become effectively screened. 
The shocks may be viewed as the boundaries of different domains of spiral or target waves, which consequently generate ``droplet'' structures in the amplitude landscape for either wave pattern.
During the initial growth process of a spatio-temporal pattern, there will also be shocks formed between the corresponding wave and its surrounding turbulent structures. 
As the droplet sizes increase, those shocks will push the surrounding defect turbulence piedmonts away. 
As an indirect result, this process accelerates the annihilation of defect pairs with opposite topological charge since the spatial regions governed by defect turbulence are being compressed, and defects with opposite charge are brought into closer vicinity. 

During the spiral nucleation process, until the entire space is filled, each spiral structure will occupy a certain domain with a length scale that is usually much larger than the wavelength of the spiral waves.
Those domains are separated by shock lines and are typically four- or five-sided polygon-like structures; their boundary shocks are approximately hyperbolic \cite{Bohr96, Bohr97}. 
Defects may also exist in the spiral far-field regime; these are thought to be passive objects that persist inside the shock region, ``enslaved'' by the vortex of a spiral domain.
This CGL (quasi-)stationary state was termed ``vortex glass'' by Huber et al. \cite{Huber92}, and ``frozen'' state by Chat\'e et al. \cite{Chate96}, since the dynamics in this regime becomes extremely slow; hence this state may persist indefinitely, at least in finite systems \cite{Aranson02}. 
Several studies have aimed at quantitatively analyzing the relaxation processes in the frozen state \cite{Braun96}, as well as the ``unlocking'' of freezing \cite{Das12, Das13, Patra13}.
Aranson et al. \cite{Aranson93a, Aranson93b} investigated the interaction between different spirals in both symmetric and asymmetric situations, as well as the mobility of spirals when driven by external white noise \cite{Sakaguchi89, Aranson98}. 
The competition between spiral and target patterns was studied by Hendrey et al. \cite{Hendrey00b}.

In this paper, we are predominantly interested in the incipient stages in the formation of spiral structures or target waves following a sudden quench from a random initial configuration, or alternatively from the defect-turbulent regime, namely nucleation processes, and the ensuing transient kinetics \cite{CGLmovie}.

\section{Numerical Scheme and Nucleation Measurement} \label{sec:scheme}

In this paper, we limit ourselves to studying the CGL on a two-dimensional spatial domain with periodic boundary conditions. 
We implement Eq.~(\ref{cglec1}) on a square lattice using the standard Euler discretization, employing central differentials in space and forward finite differential in time. 
Our discretization mesh sizes were $\Delta x = \Delta y = 1.0$, and $\Delta t = 0.001$. 
We should clarify the reason for choosing a relatively smaller (by a factor of ten compared to previous studies \cite{Huber92, Patra13}) differential time step: 
Since defect turbulence is spontaneously created by intrinsic chaotic fluctuations rather than external noise, the stability of the continuous partial differential equation (\ref{cglec1}), as well as the stability of the numerical solution scheme definitely require careful consideration. 
This caveat is supported by numerical experiments: 
The stable regime of temporal periodic solutions and phase turbulence in the $b-c$ parameter plane becomes successively compressed as the individual integration time steps are increased. 
Consequently, the differential scheme itself is rendered unstable. 
Therefore, in order to avoid such numerical artifacts and still maintain acceptable computational efficiency, we choose $\Delta t = 0.001$ as a reasonable differential time step for our implementation. 
We note that as a consistency check, we have rerun several of our simulations with a fourth-order Runge--Kutta numerical integration scheme, and obtain data and results that are fully in accord with those reported below.

In order to handle the complex field $A(\bm{x},t)$, we separate its real and imaginary parts, $A(\bm{x}, t) = A_r(\bm{x}, t) + i A_i(\bm{x}, t)$; Eq. (\ref{cglec1}) thus splits into two coupled real partial differential equations for $A_r(\bm{x}, t)$ and $A_i(\bm{x}, t)$ respectively:
\begin{eqnarray}
   \partial_t A_r &=& A_r+\nabla^2A_r-|A|^2A_r-b\nabla^2A_i+c|A|^2A_i , \nonumber \\
   \partial_t A_i &=& A_i+\nabla^2A_i-|A|^2A_i+b\nabla^2A_r-c|A|^2A_r .
   \label{cgle2}
\end{eqnarray}
The finite differential scheme described above can then be readily applied to numerically solve these two coupled partial differential equations.

To quantitatively characterize the nucleation in CGL systems, a previous study \cite{Huber92} chose to monitor the defect density, or equivalently the number of topological defects $n(t)$ in a fixed domain area $L^2$ as function of simulation time $t$. 
A corresponding typical defect separation length is then given by $l_{\rm sep}(t) = L / \sqrt{n(t)}$ for a system with linear dimension $L$.
The nucleation time can be measured by determining when the associated defect density drops below, say, two standard deviations from its statistical average in the transient turbulent state. 
However, different initial conditions induce large variations in the number and spatial distributions of defects and also, as noted in Refs.~\cite{Gil90, Wang04, Mazenko01}, the number of topological defects fluctuates around a mean value $\bar{n}$ with variance $(\Delta n)^2 = \bar{n}$. 
Thus, as in typical simulation domains $\bar{n}$ is large in the defect turbulence state, the ensuing sizeable number fluctuations combined with the strong variations induced by the initial conditions will render the thus inferred nucleation time rather inaccurate. 

\begin{figure*}[btp]
    \includegraphics[width=16cm]{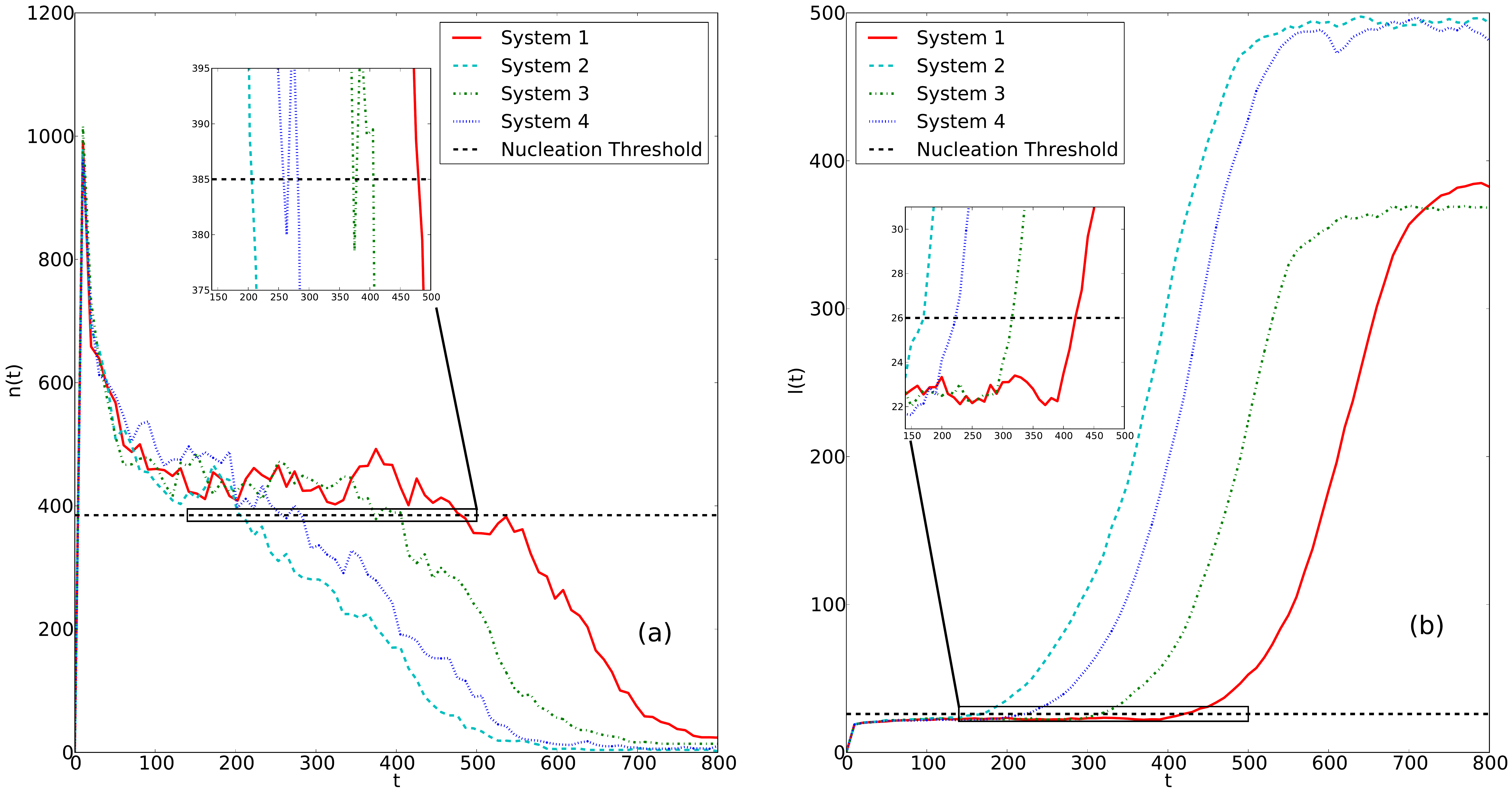} \\
    \includegraphics[width=16cm]{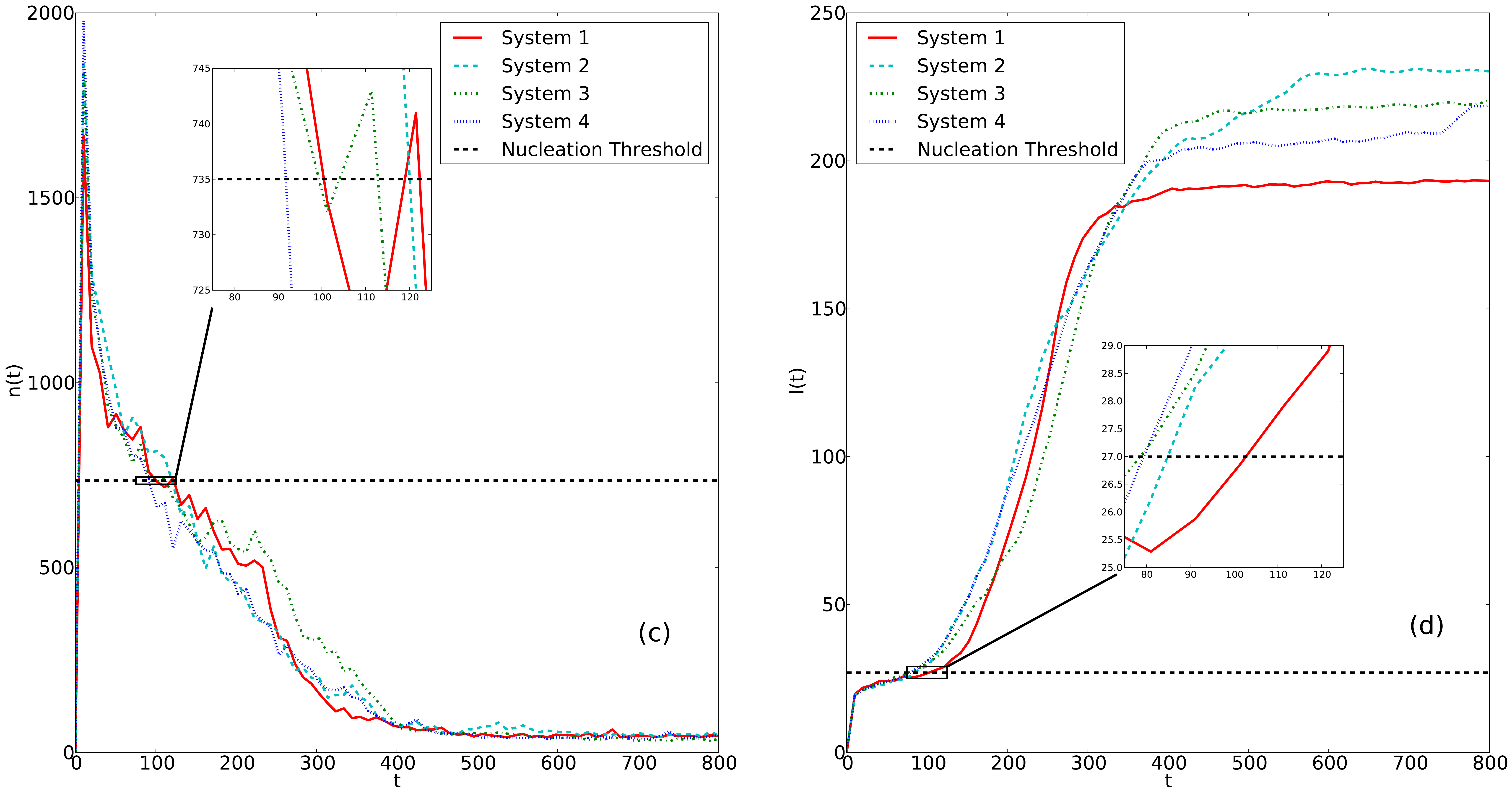}
    \caption{(a), (c) Number of topological defects $n(t)$ and (b), (d) numerically determined characteristic length scale $l(t)$, determined from the mean shock front distances, as functions of numerical simulation time $t$, for systems with control parameter pairs $b=-3.5$, $c=0.556$ (a), (b); $b=-3.5$, $c=0.44$ (c), (d). 
    The different graphs (with distinct colors) represent four independent realization runs. 
    The estimated nucleation threshold values for each plot were chosen ad hoc ``by hand and eye'': 
    (a) $n_{\rm th}=385.0$; (b) $l_{\rm th}=26.0$; (c) $n_{\rm th}=735.0$; (d) $l_{\rm th}=27.0$.}
    \label{fig:nuc_compare}
\end{figure*}
This is illustrated in Fig.~\ref{fig:nuc_compare}, where we test two individual systems with different control parameter pairs, namely (a,b): $b=-3.5$, $c=0.556$, which is close to the transition regime from defect turbulence to the frozen state; and (c,d): $b=-3.5$, $c=0.44$, which represents a location deep within the frozen region.   
In Figs.~\ref{fig:nuc_compare}(a,c), we show the measured decaying defect number $n(t)$ as function of time for four independent simulation runs, starting from different random initial configurations.
Specifically, the initial distributions of both real and imaginary parts of $A(\bm{x},0)$ are Gaussians with $\langle A_r(\bm{x}, 0) \rangle = 0 = \langle A_i(\bm{x}, 0) \rangle$ and $\langle A_r(\bm{x}, 0) A_r(\bm{x}',0) \rangle =  0.0004\,  \delta(\bm{x}-\bm{x}') = \langle A_i(\bm{x}, 0) A_i(\bm{x}',0) \rangle$, whereas $\langle A_r(\bm{x}, 0) A_i(\bm{x}',0) \rangle = 0$; hence the complex phases originate from a uniform distribution on $[0, 2 \pi)$, while the amplitudes are drawn from a Rayleigh distribution with scale parameter $\delta=0.02$.
The black dashed lines indicate the estimated nucleation thresholds, here simply determined by visual inspection of the graphs (below, we shall propose a more systematic approach to obtain this threshold and describe it via a phenomenological formula). 
Due to the large fluctuations of those individual time tracks $n(t)$, they may intersect the supposed threshold line multiple times, which induces considerable ambiguity in the measurement of associated nucleation times.
One may of course reduce this inaccuracy through running a multitude of independent numerical integrations and thus improve the statistics, yet at marked increase in computational expenses.

\begin{figure*}[btp]
    \centering
    \includegraphics[width=16cm]{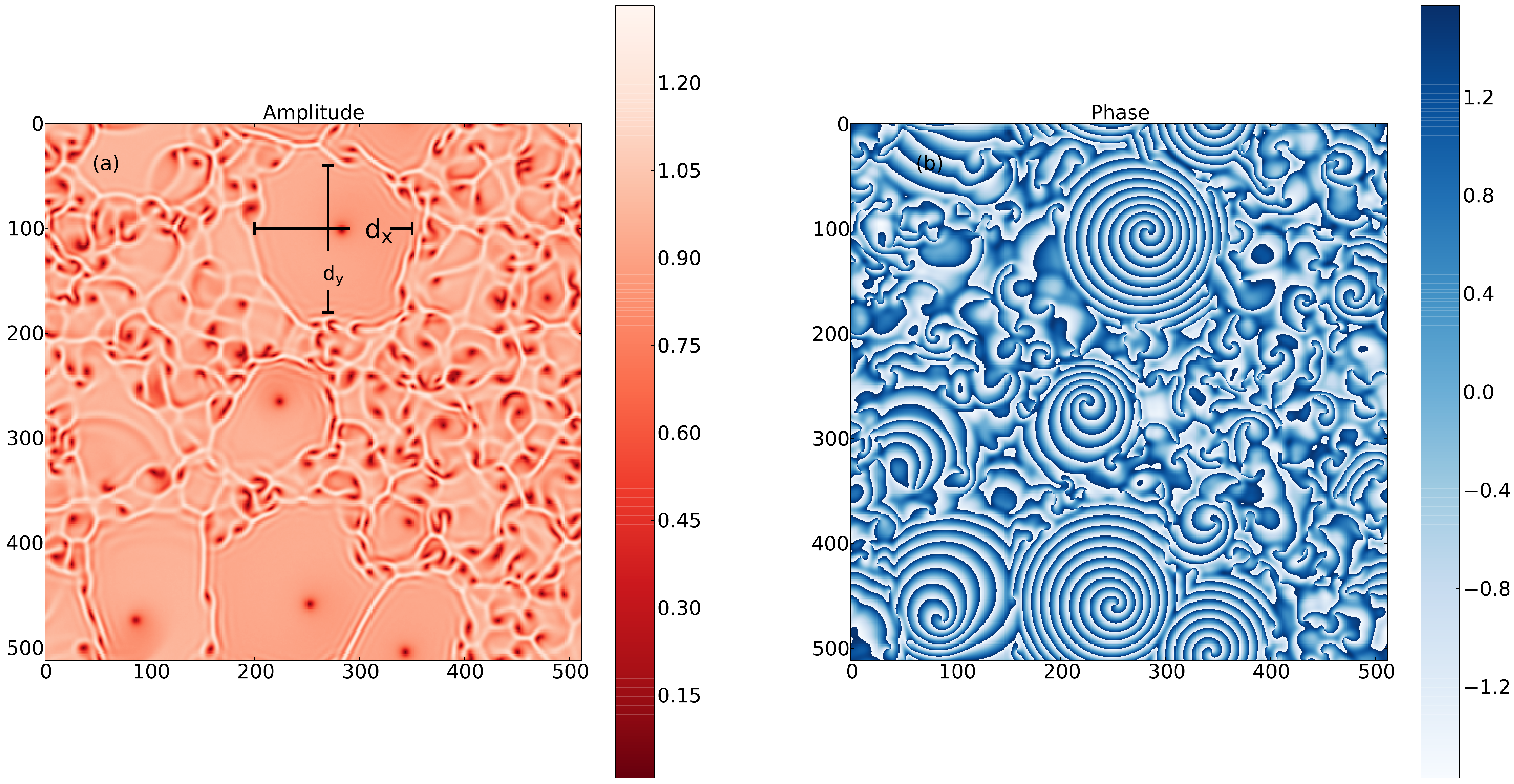}
    \caption{Amplitude (a) and phase (b) plots of the complex order parameter $A$, for control parameters $b=-3.5$, $c=0.44$. 
    The darkest points in the amplitude plot indicate the topological defects for which $|A| = 0$, while the lightest color (almost white) here indicates the shock line structures with steep amplitude gradients.
    The spiral structures are clearly visible in the phase plot within the domains separated by the shock fronts.}
    \label{fig:method0}
\end{figure*}
In order to overcome this difficulty, we propose an alternative method to extract a relevant characteristic length scale $l(t)$ directly, which is much less affected by stochastic fluctuations and consequently displays improved monotonic behavior during the nucleation process. 
To this end, we recall that the final states reached after nucleation are quasi-stationary, typically square- or pentagon-shaped single-spiral domains, which are well-separated by shocks. 
Hence we choose the average size of those domains, or equivalently, the initially growing mean distance between shock fronts, to represent a useful quantitative measure for a characteristic length scale in such CGL systems.
Specifically, we compute the distance between all pairs of adjacent shock lines both along the $x$ and $y$ directions in the amplitude plots, e.g., shown in Fig.~\ref{fig:method0}(a); their average serves as a direct measurement of the time-dependent characteristic length scale in our system. 

More details of the this method are illustrated in Fig.~\ref{fig:method0} \cite{CGLmovie}: 
We first select a point and measure both $d_x$ and $d_y$ respectively (the selected point thus is the intersection of the two depicted black lines). 
We then repeat this process by scanning all points in our square lattice and simply compute the average of all $d_x$ and $d_y$ values, yielding an estimate for the characteristic length scale.
This definition for the typical scale $l(t)$ naturally promotes the weight of large spiral nuclei, and the emergence and growth of well-established spiral structures lead to rapidly increasing values of $l(t)$; eventually, the largest spiral contributes dominantly to the thus extracted characteristic length $l(t)$.
Some preliminary measurements of this quantity for the same systems for which we determined the decaying defect number $n(t)$ are shown in Fig.~\ref{fig:nuc_compare}(b,d) as well. 
We make the following pertinent observations: First, the mean separation distances between defects $l_{\rm sep}(t)$ following from the data in graphs (a,c) roughly follow the behavior of the curves (b,d).
The data for the characteristic lengths $l(t)$ in Fig.~\ref{fig:nuc_compare}(b,d) are however clearly subject to much lower fluctuations, and from the curves' intersection with the set threshold lines allow markedly better defined estimates for nucleation times for the onset of stable spiral structures, even near the strongly fluctuating transition regime (b). 
Yet this also implies that the characteristic length scale $l(t)$ is quite sensitive to those nucleation processes, as one should expect, and a careful analysis of the entire evolution history for the complex order parameter $A(\bm{x},t)$ is required to appropriately select the proper nucleation threshold.

We shall subsequently apply the method described above to determine the growing length scales in two-dimensional CGL systems, and utilize these to further characterize nucleation processes as follows: 
First of all, we carefully investigate the evolution histories of both the phase and amplitude of the complex field $A(\bm{x},t)$ on the square lattice, and select a tentative threshold $l_{\rm th}$ according to these investigations.
Next, we collect statistically significant data, and employ this prior selected length to monitor nucleation processes: When the characteristic length $l(t) > l_{\rm th}$ exceeds the set threshold, we assume a spiral structure to have successfully nucleated. 
The spiral associated with this nucleation will then expand, and ultimately either fill the entire two-dimensional system by itself or jointly with other spiral domains. 
We hence measure the time difference from the very beginning of the numerical simulation ($t=0$) until $t=T_n$ where $l(T_n) \ge l_{\rm th}$ for the first time, and collect the resulting time interval data $T_n$ for further analysis:
In order to eliminate inaccuracies introduced by our ad-hoc choice of the nucleation threshold size $l_{\rm th}$ as well as finite-size effects, we examine the dependence of the associated dimensionless nucleation barrier on $l_{\rm th}$ and $L$ with the aid of a phenomenological formula, and finally extract an extrapolated critical nucleus size and barrier (details will be described in the following section~\ref{sec:spiral_nuc}).
We repeat these procedures and data collection with ensuing analysis for different selections of parameter pairs $(b,c)$ according to their distance to the transition line where the defect turbulence regime becomes meta-stable; these various scenarios will be discussed in the following section. 

Furthermore, since target wave patterns display quite similar shock structures as the spiral structures in the frozen state, we may also characterize the characteristic length scale of the corresponding CGL systems and their nucleation dynamics in section~\ref{sec:target_nuc} in a similar manner.

\section{Spiral Structure Nucleation} \label{sec:spiral_nuc}

We now proceed to explore spiral structure nucleation processes in two-dimensional CGL systems, for two representative different parameter pairs $(b,c)$. 
Specifically, we quench the control parameters to values that correspond to states that are situated either deep in the frozen region, or close the transition or crossover line where defect turbulence becomes meta-stable \cite{CGLmovie}. 
These two quench scenarios present distinct challenges to the subsequent statistical analysis, which hence needs to be performed carefully. 
In order to obtain robust conclusions, we gather data with sufficient statistics from many numerical integrations with different initial conditions for various system sizes $L$, which may subsequently allow us to extrapolate to the infinite-system limit. 
 
\subsection{Quench Far Beyond the Defect Turbulence Instability Line} \label{subsec:faraway}

We begin with quenches of the control parameters ($b=-3.5$, $c=0.44$) deep into the stable frozen-state regime, for which nucleation processes should happen comparatively fast, hence requiring less computation time. 
Furthermore, since the eventual frozen configurations are usually occupied by multiple spirals \cite{CGLmovie}, finite-size effects can be at least approximately eliminated through extrapolation to the $L \to \infty$ limit.

\begin{figure*}[btp]
    \centering
    \includegraphics[width=16cm]{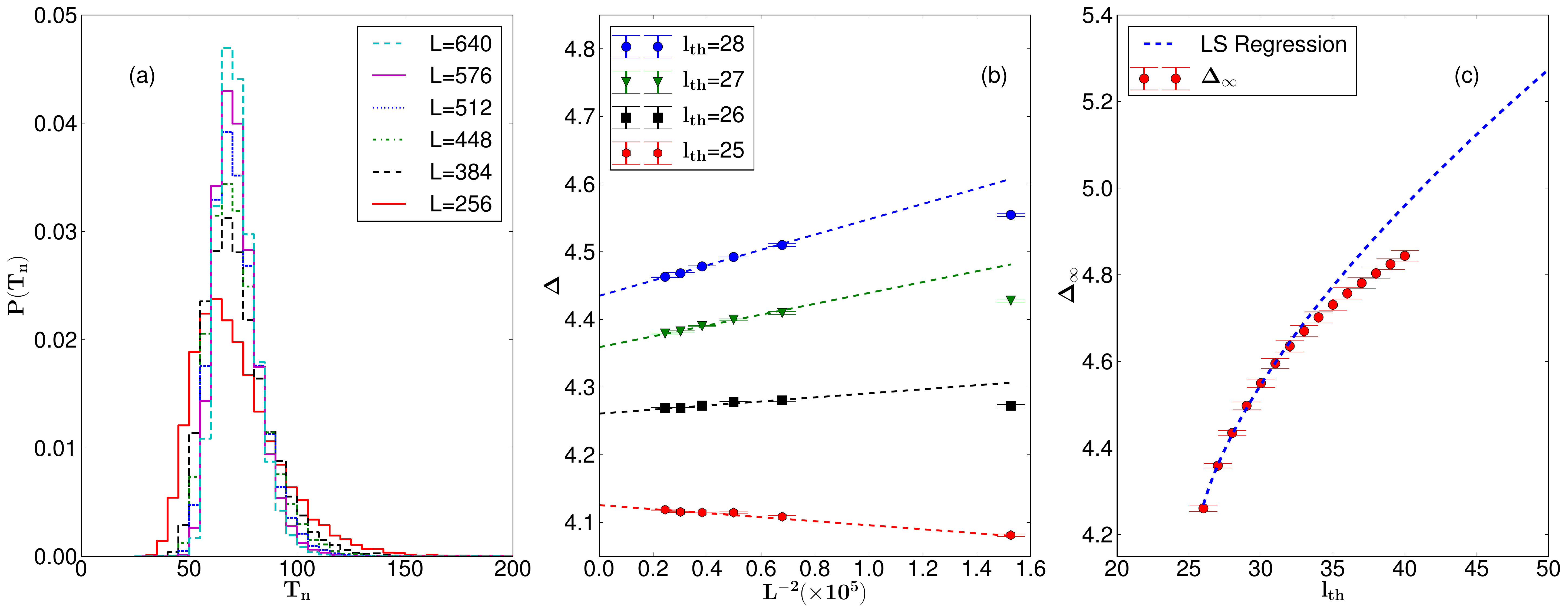}
    \caption{(a) Normalized distribution (histogram) $P(T_n)$ of measured nucleation times $T_n$ for two-dimensional CGL systems with $b=-3.5$, $c=0.44$ and different sizes (varying from $256\times256$ to $640\times640$); here we set $l_{\rm th}=27$ and ran $20,000$ realizations for each system size. 
     (b) Extracted dimensionless nucleation barrier $\Delta$ as function of the inverse system size $L^{-2}$ utilizing different values for the tentative threshold $l_{\rm th}$; the dashed lines indicate a least-square fit for the data points with the four largest system sizes. 
     (c) Infinite-size limit ($L \to \infty$) barrier $\Delta_\infty$ vs. the prior selected threshold length $l_{\rm th}$; the (blue) dashed line shows the least-square fit to Eq.~(\ref{nuc_formula1}) using the seven data points with the smallest threshold lengths.}
    \label{fig:nuc_dis}
\end{figure*}
A histogram plot for the normalized nucleation time distribution for different linear system sizes $L$ is shown in Fig.~\ref{fig:nuc_dis}(a). 
The tentative threshold length employed here is $l_{\rm th}=27$, which is simply chosen by inspection of $l(t)$ time track data, see Fig.~\ref{fig:nuc_compare}(d). 
The nucleation time distributions for these different systems look quite similar, but display steeper and sharper peaks as the system size increases, as one would expect: 
Smaller systems are more strongly affected by finite-size effects, which tend to broaden the distribution. 
Moreover, as also observed in the previous study \cite{Huber92}, we find that the total number of defects increases with growing system size, whence multiple nucleation events may happen simultaneously for larger systems.  
With increasing $L$, a slight shift of the peak to larger nucleation times is also noticeable in our data, which can be viewed as evidence for imperfect threshold length selection.
Indeed, a threshold length that renders the average nucleation time $\langle T_n \rangle$ size-invariant should constitute at least a temporary optimal choice. 
We thus collected data for multiple selected tentative threshold lengths $l_{\rm th}$ for otherwise identical system in our numerical experiment.

We proceed drawing an analogy with nucleation at first-order phase transitions in equilibrium, where thermal fluctuations may help a nucleus beyond a critical size to overcome the free-energy barrier between the metastable and stable states. 
For the CGL, we take the intensity of stochastic fluctuations in the defect turbulence regime to play the role of temperature, and also assume that nucleating stable spiral structures requires the system to overcome an effective barrier.
However, in this far-from-equilibrium system, we cannot easily quantify an effective temperature nor uniquely determine a free-energy landscape.
We may however introduce an effective dimensionless nucleation barrier $\Delta$, which we simply define via the following connection with the average transition time $\langle T_n \rangle$: 
\begin{equation}
    \langle T_n \rangle = e^\Delta 
    \label{nuc_formula0}
\end{equation}
(in units of simulation time).
Furthermore, in metastable systems finite-size effect play a crucial role, as observed in numerous numerical studies.

In the following, we describe our data analysis procedure that allows us to successfully eliminate or at least drastically mitigate finite-size corrections:
First, we propose a straightforward relation between the numerically extracted barrier from Eq.~(\ref{nuc_formula0}) and the system size $L^2$:
\begin{equation}
    \Delta \approx C_L L^{-2} + \Delta_\infty ,
    \label{nuc_formula1}
\end{equation}
i.e., we assume the finite-size corrections to scale inversely with the system size; this linear dependence of $\Delta$ on $L^{-2}$ is in fact confirmed in our data for sufficiently large $L$ in Fig.~\ref{fig:nuc_dis}(b) for four different threshold values.
Since the first term here vanishes as $L \to \infty$, $\Delta_\infty$ corresponds to the effective dimensionless barrier in the thermodynamic limit.

Second, we wish to eliminate ambiguities and uncertainties related to our ad-hoc choice of the threshold length $l_{\rm th}$ used to measure the nucleation times.
Indeed, the extrapolated value for  $\Delta_\infty$ does depend significantly on the selection of $l_{\rm th}$, as is evident in Fig.~\ref{fig:nuc_dis}(b): 
A larger threshold length implies bigger nucleation droplet size, the formation of which definitely requires longer time. 
Yet we may immediately rule out some choices first based on Eq.~(\ref{nuc_formula1}); for example, when $l_{\rm th}=25$, we find a negative slope ($C_L < 0$) which contradicts the basic assertion that activation barriers should become entropically reduced in larger systems; therefore, we may consider $l_{\rm th}=25$ an inappropriate choice.
Indeed, since our goal is to also reduce finite-size effects inasmuch as at all feasible, it appears natural to pick a threshold length that will provide us with a size-invariant mean nucleation time; this heuristic criterion effectively eliminates the strongest finite-size corrections induced by different choices for $l_{\rm th}$.
For the data displayed in Fig.~\ref{fig:nuc_dis}(b), we infer that this optimal threshold $l_c$ should lie in the range $(25.0, 26.0)$, but closer to the upper bound, according to the slope of the four dashed lines. 
One might thus iterate the above simulation steps to further confine the interval for an optimized $l_c$; however, this procedure would be computationally quite expensive, and also impeded by sizeable statistical errors that render distinctions between almost flat functions $\Delta(L^{-2})$ very difficult.
  
We therefore propose an alternative approach in postulating a purely phenomenological formula to model the relationship between the extrapolated $\Delta_\infty$ and $l_{\rm th}$:
\begin{equation}
    \Delta_\infty(l_{\rm th}) = \Delta_0 + \Delta_1 \left( l_{\rm th} - l_c \right)^\theta \ ,
    \label{barrier0}
\end{equation}
which empirically fits our data quite well for $l_{\rm th} \approx l_c$, as shown in Fig.~\ref{fig:nuc_dis}(c).
Naturally, we would expect Eq.~(\ref{barrier0}) to hold only when $l_{\rm th} \approx l_c$, and consequently restrict our subsequent data analysis to this regime where the results are nearly independent of the originally selected threshold length $l_{\rm th}$.
We note that selecting a too large value for $l_{\rm th}$ would bias configurations towards super-critical nuclei, which essentially have already reached the frozen state. 
Similarly, when estimating our error bars for, e.g., the least-square fits to our empirical formula (\ref{barrier0}), we give more weight to the data points closer to $l_c$.

\begin{figure*}[btp]
    \centering
    \includegraphics[width=16cm]{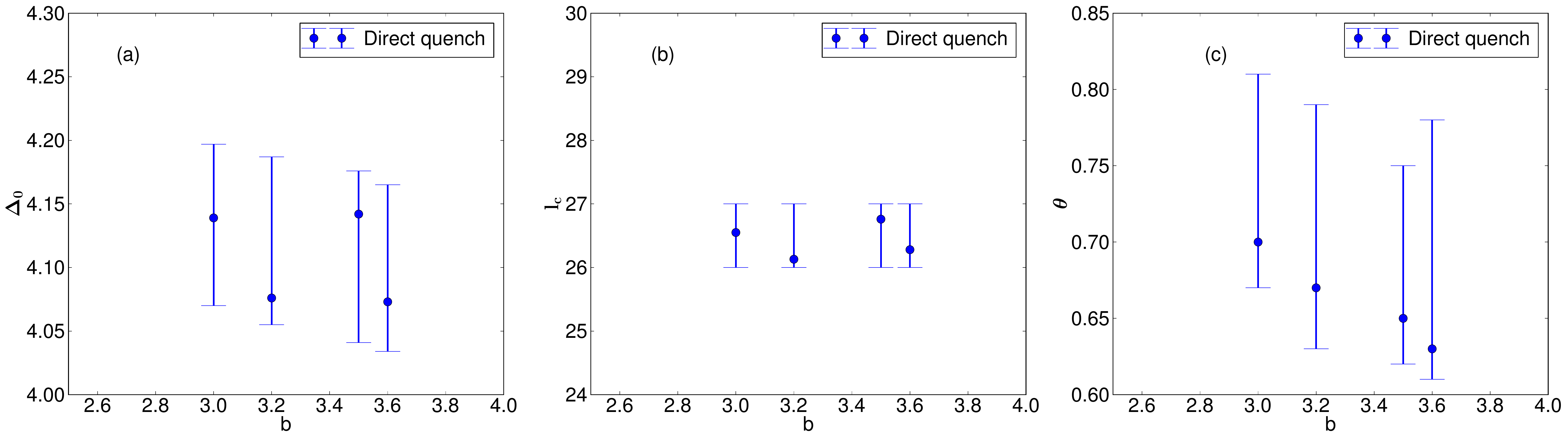}
    \caption{(a) Size-invariant dimensionless barrier $\Delta_0$, (b) critical threshold length $l_c$, and (c) exponent $\theta$ as functions of the control parameter $b$, with $c = -0.40$ held fixed.
	These quantities are extracted by fitting our numerical data to the empirical formula (\ref{barrier0}); the label ``Direct quench'' indicates that the CGL systems here are directly quenched from random initial configurations into the frozen regime.}
    \label{fig:nuc_imb0}
\end{figure*}
\begin{figure*}[btp]
    \centering
    \includegraphics[width=16cm]{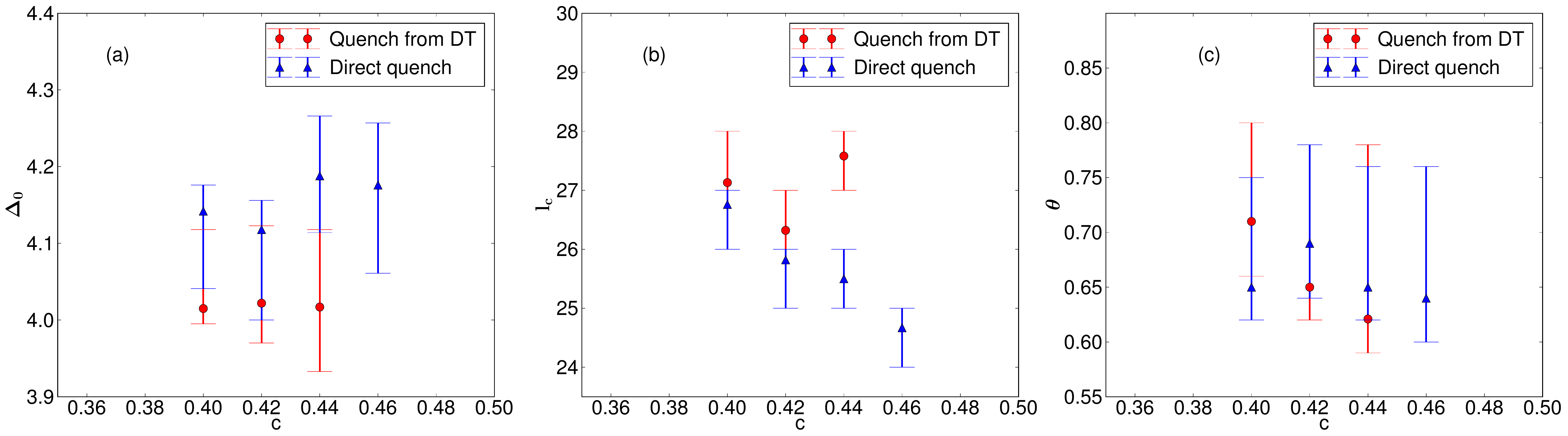}
    \caption{(a) Size-invariant dimensionless barrier $\Delta_0$, (b) critical threshold length $l_c$, and (c) exponent $\theta$ as functions of the control parameter $c$, with $b = -3.50$ held fixed.
	The label ``Quench from DT'' (data plotted in red) indicates that these systems were initialized in the defect turbulence regime with $b=-3.4$ and $c=1.0$, remained in this phase for a simulation time interval $\Delta t = 50$, and were subsequently quenched into a frozen state characterized by the parameter pairs $(b,c)$ listed.}
    \label{fig:nuc_imc0}
\end{figure*}
We have applied the data analysis method described above to investigate representative CGL parameter pairs $(b,c)$. 
In order to check for systematic dependencies, we hold either $b$ or $c$ fixed, and vary the other control parameter, to yield two complementary data sets. 
Since we investigate the focusing quadrants with $b c < 0$, we may utilize the fundamental gauge symmetry $(A,b,c) \to (A^*,-b,-c)$ to restrict ourselves to fixing either $b < 0$ with varying $c > 0$, or vice versa.
Figures~\ref{fig:nuc_imb0} (for fixed $c = -0.4$) and \ref{fig:nuc_imc0} (with fixed $b = - 3.5$) show our results for $\Delta_0$, $l_c$, and $\theta$ in Eq.~(\ref{barrier0}) for each parameter set. 
In both sets of parameters, upon increasing $c > 0$ or $b > 0$, respectively, the system approaches the transition line beyond which persistent defect turbulence is stabilized \cite{Aranson02}.

We first observe that the extracted dimensionless nucleation barriers $\Delta_0$, which according to our analysis should be essentially independent of the choice of the threshold length and effectively represent values extrapolated to infinite system size, appear to display no statistically significant dependence either on our adjusted sets of parameter values $b$ or $c$, but merely fluctuate around $\Delta_0 \approx 4.15$. 
This robust positive value is likely a consequence of the fact that those parameter pairs reside far away from the crossover limit. 
Hence the meta-stable transient turbulent state observed in Ref.~\cite{Huber92} does not become apparent, and our systems directly nucleate from the imposed fully random initial configurations to persistent spiral structures in the frozen state with typical nucleation times that do not differ markedly for the parameter range tested here.
Thus, within the parameter ranges explored in this work, it appears that the effective extrapolated nucleation barrier $\Delta_0$ remains positive, and based on the absence of any noticeable decrease upon approaching the instability line of the defect turbulence state, we conjecture that it remains non-zero throughout the frozen (or ``vortex glass'') state. 
This would imply that the transition from defect turbulence to the frozen state is discontinuous.
Indeed, in our data in Fig.~\ref{fig:nuc_imc0} for fixed $b=-3.5$, one might even discern that $\Delta_0$ rather slightly grows with increasing $c$, which may indicate an enhanced effective surface tension.
However, fluctuations also become larger upon approaching the transition limit, and we would need much improved statistics to make a definitive conclusion.

Second, we find the extracted threshold lengths $l_c$ that consistently monitor the nucleation for different system sizes to slightly drop with increasing $c>0$ for fixed $b < 0$, see Fig.~\ref{fig:nuc_imc0}, while they remain constant within our error bars as function of $b > 0$ for fixed $c < 0$, Fig.~\ref{fig:nuc_imb0}. 
This observation may be explained through the fact that upon varying the parameter $c$ from $c=0.4$ to $c=0.46$ at constant $b=-3.5$, one approaches the transition regime faster, and comes closer to it, than for the explored $b$ range at constant $c=-0.40$.
Third, we do not discern significant systematic variations of the fit exponent $\theta$ in Eq.~(\ref{barrier0}) with $b$ or $c$, see Figs.~\ref{fig:nuc_imb0}(a) and \ref{fig:nuc_imc0}(a); within our numerical accuracy, this exponent appears universal for the frozen states with the distinct parameter pairs investigated here.

As mentioned above, our results for the spiral structure nucleation processes appear at variance with the findings in Ref.~\cite{Huber92}. 
The main reason for this discrepancy is that we consider quenches of our system much deeper into the frozen regime, whence we do not observe meta-stable defect turbulence states. 
Our data consequently do not allow us to decouple the initial decay process, suggested by Huber et al., from the actual nucleation process that overcomes the barrier.
For the cases studied here, the transient relaxation kinetics is obviously fast, and we may therefore simplify our nucleation time measurements by simply combining these two processes.
In order to check if this simplification is adequate, we have applied a different quench protocol to investigate nucleation into frozen spiral structures:
Namely, instead of directly adjusting the parameter pair $(b,c)$ to the frozen state regime, we first quench the system to the defect turbulence region.
We subsequently run its dynamics sufficiently long for the system to reach a stationary (quasi-stable) turbulent state with roughly constant density, and only then quench the system again into the frozen state regime. 
Thus we first explicitly set the system up in the defect turbulence region, and later let the spiral structures nucleate from this configuration.

We apply the same techniques as previously to analyze spiral structure nucleation processes in this scenario, except that the nucleation time is now measured as the elapsed interval after performing the second quench.
We choose the corresponding control parameter pair for the explicit defect turbulence state uniformly as $b=-3.4$ and $c=1.0$. 
Yet there is of course still stochastic variability for the ensuing nucleation processes; hence we test three pairs of control parameters, namely $(b,c)=(-3.5,0.4)$, $(-3.5,0.42)$, and $(-3.5,0.44)$.
Inspection of the resulting data yields that the explicit defect turbulence intermediate state merely causes an overall shift of the nucleation time distribution towards lower values; and this offset turns out small compared with the average nucleation time. 
The associated size-invariant nucleation barriers $\Delta_0$, the critical threshold lengths $l_c$, and fit exponents $\theta$ are shown (as red data points) in Fig.~\ref{fig:nuc_imc0}.
The thus extracted dimensionless nucleation barrier values are consistent with the direct quench results within our error bars.
This suggests quite similar nucleation kinetics starting from either random initial configurations or from a pre-arranged defect turbulence state. 
Therefore, our discontinuous transition conclusion should hold even if one explicitly decouples the initial decay from random initial configurations from the subsequent relaxation from the defect turbulence state, at least in this deep-quench scenario.
We do however observe deviations in the measured optimal critical threshold lengths in Fig.~\ref{fig:nuc_imc0}(b) for these two quench strategies, with the data obtained from the case with intermediate defect turbulence states resulting in slightly larger and more uniform values for $l_c$.
Apparently, these minor variations in the critical threshold $l_c$ do not strongly affect the extracted size-invariant barriers $\Delta_0 \approx 4$, which show no tendency to decrease upon approaching the instability line.
Since the line of metastability runs essentially parallel to the $b$ axis in the $(b,c)$ parameter space near the point $(-3.5,0.4)$, we expect a very weak dependence of $l_c$ and $\Delta_0$ on $c$ for quenches from the defect turbulence state as well, similar to the results shown in Fig.~\ref{fig:nuc_imb0}.

\subsection{Quench Close To the Instability Line} \label{subsec:near}

\begin{figure*}[btp]
    \centering
    \includegraphics[width=16cm]{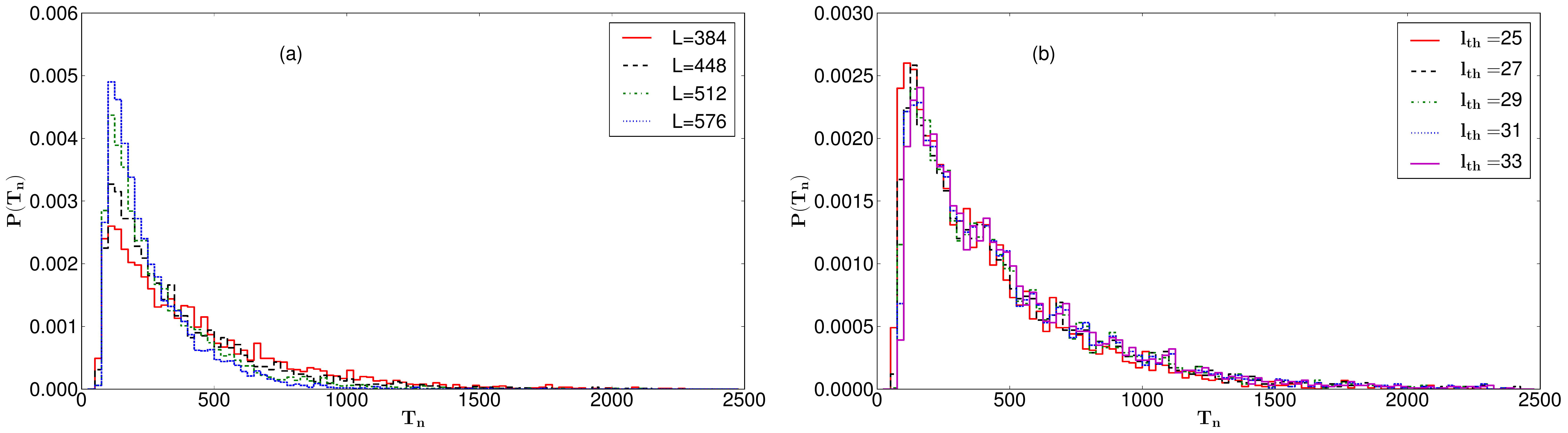}
    \caption{Normalized nucleation time distributions $P(T_n)$ for two-dimensional CGL systems with $b=-3.5$ and $c=0.556$. 
    (a) Data for varying the system size $L^2$ from $384\times384$ to $576\times576$, with ad-hoc selected nucleation threshold length $l_{\rm th}=25$. 
    (b) Histograms for different nucleation thresholds $l_{\rm th}$ at fixed system size $L = 384$; $4,000$ independent realizations were run until simulation time $t = 2400$ for each histogram.}
    \label{fig:nuc_dis1}
\end{figure*}
Next we proceed to investigate a parameter quench close to the transition line \cite{CGLmovie}, applying similar numerical methods. 
When approaching the crossover regime, as shown in Fig.~\ref{fig:nuc_compare}(a,b), a meta-stable state with a well-defined topological defect density and corresponding characteristic droplet size is formed. 
However, the proximity to the instability regime in the $b$-$c$ parameter space induces much larger fluctuations than in the deep-quench scenario of the previous subsection.
Therefore, a wider nucleation time distribution with larger mean is to be expected in this case. 
Indeed, this prediction is borne out by our simulations, for which we gathered data for quenches to the control parameter values $b=-3.5$ and $c=0.556$, exactly the same as used in Fig.~\ref{fig:nuc_compare}, with comparatively lower statistics ($4,000$ realizations were run for each distribution).
The resulting normalized nucleation time histograms are displayed in Fig.~\ref{fig:nuc_dis1} for various system sizes $L$ (a) and different choices for the threshold length $l_{\rm th}$ (b).

\begin{figure*}[btp]
    \centering
    \includegraphics[width=16cm]{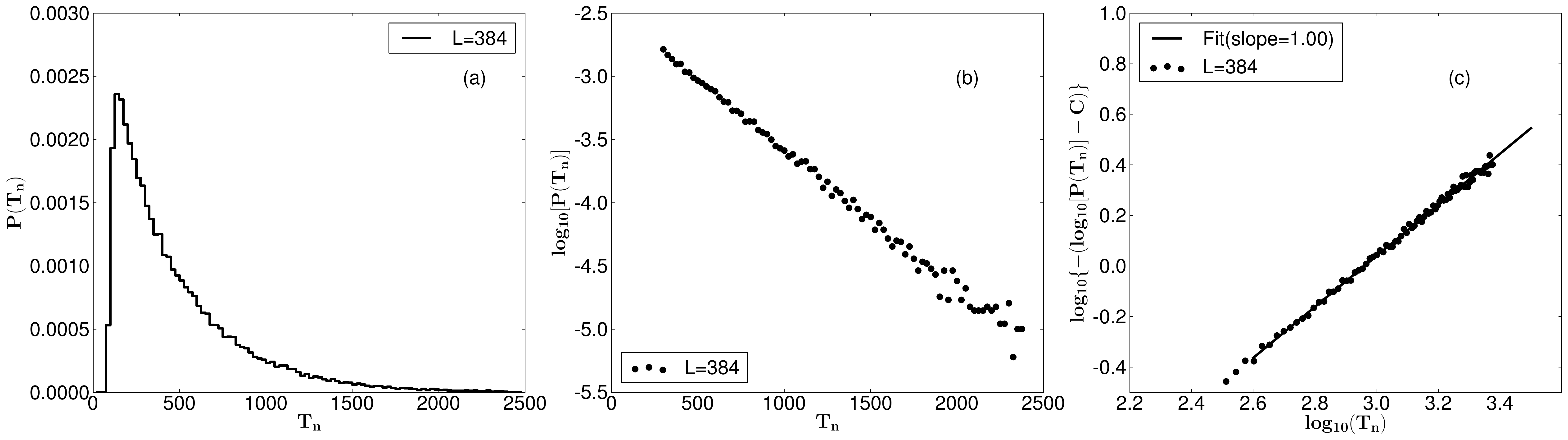}
    \caption{Normalized nucleation time distribution $P(T_n)$ in linear (a) and logarithmic (b) scale for two-dimensional CGL systems with $b=-3.5$, $c=0.556$, linear system size $L = 384$, and nucleation threshold length $l_{\rm th}=32$; 
    (c) shows a double-logarithmic plot of the data in (b) with constant offset $C$.
    $40,000$ independent realizations were run until simulation time $t = 2400$.}
    \label{fig:nuc_dis2}
\end{figure*}

We note that no significant difference is discernible between the nucleation time histograms in Fig.~\ref{fig:nuc_dis1}(a) for the data obtained with linear system sizes $L = 384$ up to $576$, especially in the long-time tails of the distributions, in contrast with the results shown in Fig.~\ref{fig:nuc_dis} for quenches far beyond the transition line.
The reason for this distinct behavior is, perhaps counter-intuitively, a larger finite-size effect in this present case: 
Indeed, most of the CGL systems in domains of size $384\times384$ or $448\times448$ will actually in the end be occupied by a single large spiral, owing to stronger fluctuations and lower nucleation probabilities.
Therefore, the ensuing nucleation distributions are quite similar, while the two bigger simulation domains may accommodate perhaps two or three spiral structures, resulting in slightly narrower and sharper peaks in the associated nucleation time distributions.
These observations unfortunately render the extrapolation method detailed in the previous subsection impractical in the present situation, as this would require runs on prohibitively much larger systems which exceeds our current computational resources.
Furthermore, the rather small differences between the nucleation time distributions obtained for different pre-set threshold lengths visible in Fig.~\ref{fig:nuc_dis1}(b) do not allow systematic studies of the dependence on $l_{\rm th}$ either.
Consequently we focus on characterizing the long, ``fat'' tail in the nucleation time distributions instead, which were in fact comparatively insignificant in the deep-quench scenario.
Hence we now restrict ourselves on relatively small systems with $L=384$, for which the final quasi-stationary state most likely contains a single spiral, but measure the nucleation time distribution for fixed size and a single set nucleation threshold $l_{\rm th}=32$ with the same control parameters, but with much better statistics, namely running $40,000$ independent realizations; the resulting histogram is depicted in Fig.~\ref{fig:nuc_dis2}.
As shown in Fig.~\ref{fig:nuc_dis2}(b), the decaying part of this nucleation time distribution is well-described by a simple exponential form
\begin{equation}
    P(T_n) \propto e^{- T_n / \tau} ,
    \label{ex_dis}
\end{equation}
as is further confirmed by the double-logarithmic fit for the long-time ``fat'' tail in Fig.~\ref{fig:nuc_dis2}(c); i.e., our measured nucleation time can be viewed as an exponentially distributed random variable in the long tail regime.
We point out that a nucleation study in the meta-stable region for a two-dimensional ferromagnetic Ising spin system with Glauber dynamics in systems subject to a polarizing magnetic field at zero temperature yielded very similar results \cite{Neves91}. 
For that model, the normalized nucleation time of a single rectangular droplet of positive spins in a bulk region of opposite alignment has been proven to represent be a unit-mean exponentially distributed random variable.
This fact in turn is a consequence of the discontinuous nature of the phase transition separating both ferromagnetic configurations below the critical temperature upon tuning the magnetic field. 
The average lifetime of the metastable state is also rigorously shown to satisfy Eq.~(\ref{nuc_formula0}) with appropriate free-energy barrier divided by $k_{\rm B} T$ \cite{Park04}, consistent with our conclusion in subsection~\ref{subsec:faraway}.

It should however be noted that our CGL nucleation processes are not completely captured by this equilibrium asymptotic analysis, and our measured nucleation time distributions are not perfectly exponential, in neither of the quench scenarios discussed above.
This is due to several facts: 
First of all, for systems that are quenched from random initial configuration, there exists a finite initial decay interval until they attain a meta-stable state. 
This was pointed out by Huber et al. \cite{Huber92}, who consequently applied an overall time shift for the nucleation histograms.
Second, finite-size effects clearly affect numerically obtained nucleation time distributions. 
As is apparent in Fig.~\ref{fig:nuc_dis1}(a), the function $P(T_n)$ approaches an exponential shape for larger systems, and its rising flank tends to disappear as $L \to \infty$.
For quenches close to the instability line, computational limitations prevent us from systematically eliminating finite-size effects. 
On the other hand, CGL configurations quenched far beyond the crossover regime are characterized by separate spiral droplet domains that are separated by sharp shock fronts.
These structures are very stable, at least in simulations with finite system sizes, and hence effectively prevent the merging of individual spirals to form increasingly large droplets.
Consequently, the CGL systems indeed become ``frozen'', in stark contrast with the slow coarsening kinetics in the ferromagnetic Ising spin system.

\section{Target Wave Nucleation} \label{sec:target_nuc}

Upon introducing an appropriate localized inhomogeneity, one may trigger target wave patterns for the CGL \cite{CGLmovie}. 
These small inhomogeneous spatial regions then play the role of pacemakers, which locally alter the oscillation frequency \cite{Vanag01}. 
Another useful analogy is to view the small inhomogeneous regime as a slit, whence target wave patterns can be interpreted as diffraction phenomena. 
Thus the spatial extend of the inhomogenity should be comparable to the wave length of the resulting target structure. 
In contrast to the spiral waves discussed in the previous section, target waves carry zero topological charge $n=0$. 

\begin{figure*}[btp]
    \centering
    \includegraphics[width=16cm]{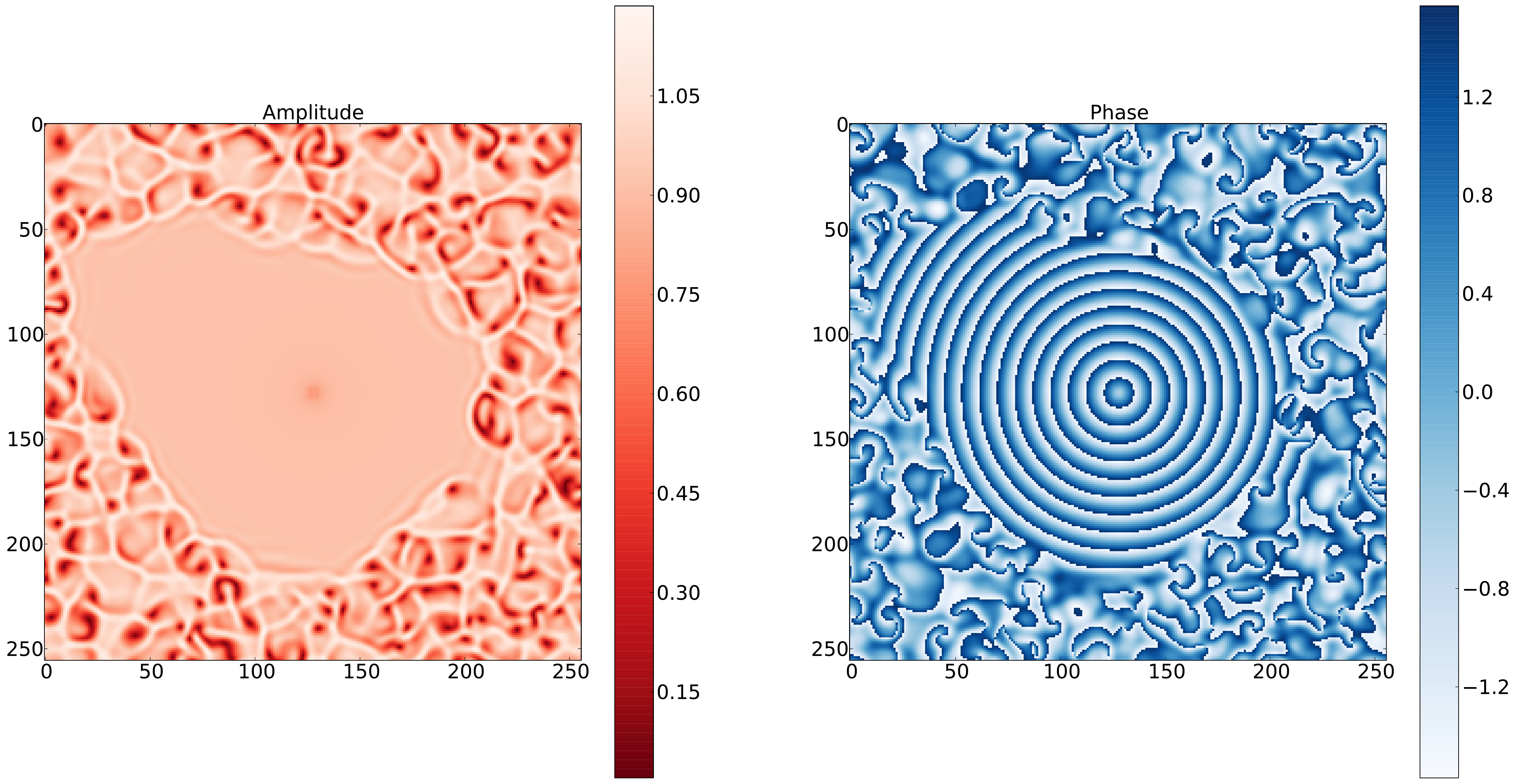}
    \caption{Amplitude (a) and phase (b) plots of the complex order parameter $A$, with bulk control parameters $b=-1.4$, $c=0.9$ in a $256 \times 256$ system; in the central $4\times 4$ block, instead $b=-1.4$, $c=0.6$ (discernible as the small square structure with different coloring in the center). 
    The configuration is shown at numerical time $t=1200$; in (a), the lightest color (almost white) indicates shock line structures with steep amplitude gradients.}
    \label{fig:target_ex}
\end{figure*}
In the following, we explore transient kinetics in the formation of target waves, and indeed characterize it in terms of nucleation dynamics, applying the same techniques as detailed above for incipient spiral structures.
To this end, we set our CGL simulations up as follows: 
We initiate them with fully randomized configurations, and select control parameters according to previous work that aimed to control defect turbulence by means of target waves \cite{Jiang04}, namely we fix $b=-1.4$ over the whole system, and set $c=0.9$ to obtain an absolutely unstable defect turbulence configuration, except for a ``central'' patch of $4 \times 4$ sites, where $c=0.6$, which corresponds to a stable frozen state.
We remark in passing that other inhomogeneities may also generate target waves, as investigated in Ref.~\cite{Hendrey00a}. 
There, however, the basic Eq.~(\ref{cglec1}) becomes modified by an addition to the imaginary part in the coefficient of the linear term which causes an explicit frequency shift. 
In this case, there exist two different types of target wave solution; we will however not consider this type of inhomogeneity as the ensuing phenomenology cannot be directly compared with our spiral nucleation study.

With the setup described above, a target wave oscillating with the pacemaker's frequency becomes spontaneously excited, after sufficient computation time has elapsed.
Furthermore, the number of target wave structures is determined by how many distinct impurity regions are introduced. 
An example for a resulting coherent structure is depicted in Fig.~\ref{fig:target_ex}, for which a single local inhomogeneity was implanted, whence exactly one emergent target wave is expected.
The existence of the shock-line structures separating the near-circular target waves from the surrounding defect turbulent state ensure that our method to monitor nucleation events using a characteristic growing length scale $l(t)$ is applicable here as well. 
However, close observation of the temporal evolution of this target wave structure invalidates our strategy of section~\ref{subsec:faraway} to extract an invariant threshold length. 
Due to the significant spatial asymmetry for the target wave expansion into its turbulent environment, there is a fair chance that the initiated droplet structure may suddenly shrink and subsequently attempt to reform towards a more symmetric droplet shape.
This leads to significant non-monotonic variations in the characteristic droplet size, and concomitant ambiguity in determining any incipient nucleation event.  
We note that this phenomenon may be associated with the fact that the inhomogeneity is introduced by hand, as well as the imposed periodic boundary conditions. 
Furthermore, the nucleation of target wave droplets is markedly more difficult than that of spiral structures as investigated in section~\ref{sec:spiral_nuc}, since there is now only one pre-assigned nucleation center, and the surroundings, which reside in a stable defect turbulence state, are subject to stronger fluctuations.
Consequently we proceed to tentatively investigate target wave nucleation processes with a single ad-hoc selected nucleation threshold, under the assumption that this will suffice to characterize the system's time evolution, similar as in section~\ref{subsec:near} for spiral structures in the near-instability quench scenario. 

\begin{figure*}[btp]
    \centering
    \includegraphics[width=16cm]{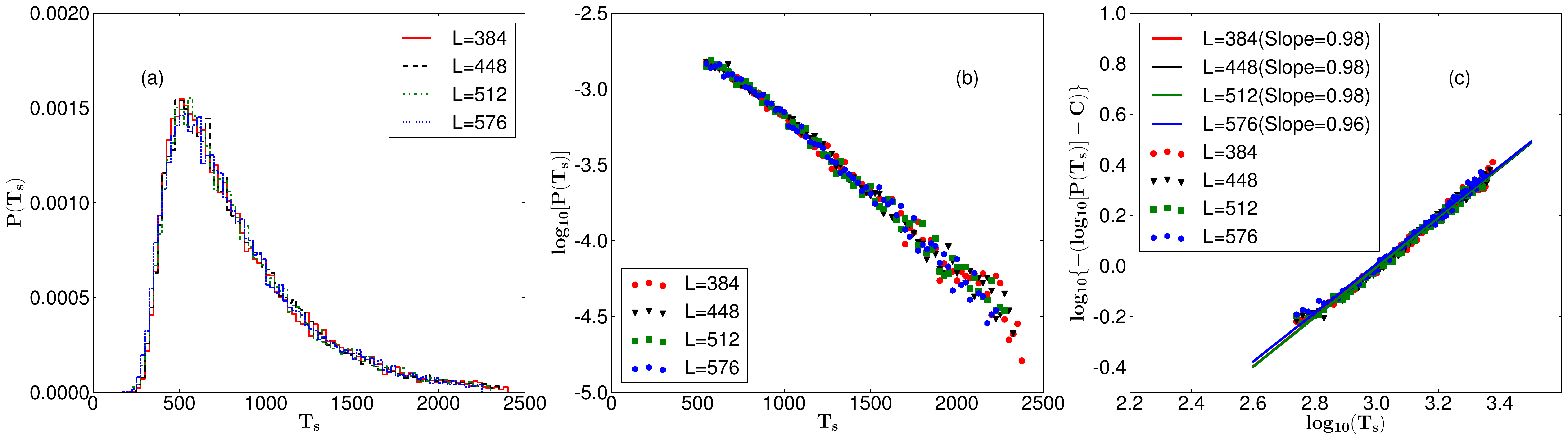}
    \caption{Normalized nucleation time distribution $P(T_s)$ in linear (a) and logarithmic (b) scale for two-dimensional CGL systems with bulk control parameter values set to $b=-1.4$ and $c=0.9$, whereas $c=0.6$ in the central $4 \times 4$ patch, for varying linear system size ranging from $L=384$ to $576$. 
    $T_s$ represents the measured nucleation time adjusted by a size-dependent shift in order to achieve data collapse; we set the nucleation threshold length to $l_{\rm th}=17$, and ran $20,000$ independent realizations for each system size; 
   (c) shows a double-logarithmic plot of the data in (b) with constant offset $C$.}
    \label{fig:target_dis}
\end{figure*}
Our measured nucleation time distributions, obtained for different system sizes with $20,000$ runs for each $L$, are shown in Fig.~\ref{fig:target_dis}.
Here, $l_{\rm th}=17$ constitutes an appropriate choice of the threshold length; according to our observations this value is just a bit larger than the optimal critical threshold.
As anticipated, the typical nucleation times for the target waves are much longer than those for spiral structures. 
In fact, some simulation runs did not lead to successful nucleation events by the time $t=2,400$ when our runs were terminated. 
Hence, the actual number of realizations over which the data were averaged are actually different for each system size, as listed in Table~\ref{tab:nuc_number}. 
Upon increasing $L$, we observed a manifest shift to higher values in the directly measured nucleation times, which reflects the enhanced stability of the bulk defect turbulence regime with growing total system size at fixed spatial extent of the central nucleation inhomogeneity.
Correspondingly, we detect a slight decrease in the number of successful nucleation incidents (Table~\ref{tab:nuc_number}). 
In order to facilitate the comparison of data resulting from the different system sizes, we have shifted the nucleation time histograms by hand to collapse them on top of each other. 
Aside from that overall shift, these distributions (as functions of the shifted nucleation times $T_s$) are quite similar because of the identical nucleation centers. 
Furthermore, they display long, fat tails in the large-time regime, akin to spiral droplet nucleation for quenches close to the transition line in parameter space. 
Indeed, as in that quench scenario, the CGL systems dominated by target waves are ultimately also occupied by a single droplet structures, since of course only one inhomogeneous nucleus was placed in the simulation domain. 
The large-$T_s$ tails are again of an exponential functional form, as demonstrated by the logarithmic and double-logarithmic data plots in Fig.~\ref{fig:target_dis}.
We note that this conclusion holds even if larger values of $l_{\rm th}$ are utilized.
This remarkable consistency between spiral structure and target wave nucleation may be explained in a natural manner by simply viewing the latter as externally stabilized structures with topological charge $n=0$ \cite{Hendrey00a}, akin to spiral waves with $n = \pm 1$, whence their nucleation events display similar exponential statistics. 
\begin{table}[ht]
     \centering
     \begin{tabular}{|c|c|}
        \hline
        System size & Number of nucleated systems \\
        \hline
        $384\times384$ & $19,792$ \\
        \hline
        $448\times448$ & $19,733$ \\
        \hline
        $512\times512$ & $19,732$ \\
        \hline
        $576\times576$ & $19,589$ \\
        \hline
     \end{tabular}
     \caption{Number of systems that have nucleated successfully by computation time $t=2,400$ among $20,000$ independent realizations for different system sizes.}
     \label{tab:nuc_number}
\end{table}

\section{Conclusions} \label{sec:conclusions}

We have studied the transient dynamics from the defect turbulence to the frozen state in the two-dimensional CGL. 
To this end, we have numerically solved the CGL equation explicitly on a square lattice, and investigated the associated nucleation process of stable spiral as well as target wave structures, which eventually dominate the whole two-dimensional domain when the quasi-stationary frozen state is reached \cite{CGLmovie}.
In order to quantitatively and reliably characterize the nucleation kinetics, we have proposed a computational method to systematically extract the characteristic nucleation lengths for various parameters in the CGL systems, which signify the typical size of incipient droplet structures in the amplitude field of the complex order parameter. 
We have collected sufficient data and for various system sizes to ensure decent statistics, and in the deep-quench scenario allow for extrapolation to infinite system size.

For the spiral droplet nucleation study, we prepare our system with random initial configurations and quench it to the meta-stable defect turbulence regime in control parameter $(b,c)$ space, in two scenarios either near and far away from the transition or crossover line to the frozen region. 
By means of our extrapolation method and proposed phenomenological formula to eliminate artifacts related to the choice for the nucleation threshold as well as finite-size effects, we have extracted a finite effective dimensionless nucleation barrier for CGL systems that are quenched far away from the instability line.
We posit that this non-zero, and approximately constant nucleation barrier indicates a discontinuous transition from the defect turbulence to the frozen state displaying persistent quasi-stationary spiral structures.
We have found indications that this conclusion holds for various quenches into the defect turbulence region with different system control parameters, and it appears that the fit exponent $\theta$ in Eq.~(\ref{barrier0}) might be universal as well.
In addition, we have considered a distinct quench scenario with an intermediate explicit turbulent state to evaluate the effect of quite different initial conditions on the ultimate spiral nucleation processes. 
We have detected only minor differences in both critical nucleus sizes and effective nucleation barriers between both situations, apparently confirming a robust discontinuous transition picture.
 
On the other hand, for quenches to regions located near the transition line in parameter space, we obtain an exponential decay in the measured nucleation time distributions, with long ``fat'' tail.
This finding is remarkably similar to spin droplet nucleation in ferromagnetic spin systems with non-conserved Glauber dynamics in finite two-dimensional lattices with periodic boundary conditions subject to a polarizing external field in the zero-temperature limit.
Drawing this analogy provides us with additional evidence for our discontinuous transition conclusion, despite obvious differences between nucleation processes in two-dimensional non-equilibrium CGL systems and such equilibrium ferromagnetic lattices. 
The results from our combined two different quench scenarios reinforce our conclusion that the transition between the defect turbulence and spiral frozen states is likely discontinuous.

Finally, we have investigated nucleation processes for different patterns that can be also observed in some experimental systems, namely target waves. 
In this situation, in the eventual frozen state our systems become filled with target wave rather than spiral structures in the phase map of the order parameter, which also are droplet-like in the associated amplitude field. 
To trigger the nucleation process for target waves, one needs to introduce some specific inhomogeneity into the two-dimensional CGL systems. 
(We note that there also exist other methods to generate target wave structures \cite{Li10} which are beyond the scope of this paper.)
 
We have applied the same analysis method as for our spiral-wave nucleation study in the near-transition line quench case, and arrive at the remarkable conclusion that the nucleation kinetics in these two very distinct situations, leading to rather different final wave structures, are in fact quite similar:
Again we observe an exponential distribution for target wave nucleation, accompanied by the characteristic long fat tails that are indicative of an associated discontinuous transition from the defect turbulence to the frozen state.
We hope that similar methods will be utilized in the future, perhaps with improved and more powerful computational resources, to quantitatively explore and robustly characterize nucleation features for both the CGL as well as other stochastic non-linear dynamical systems.

\begin{acknowledgments}
The authors are indebted to Bart Brown, Harshwardhan Chaturvedi, Michel Pleimling, and Priyanka for helpful discussions, and specifically wish to thank the latter for a careful critical reading of the manuscript draft.
This research is supported by the U.S. Department of Energy, Office of Basic Energy Sciences, Division of Materials Science and Engineering under Award DE-SC0002308.
\end{acknowledgments}


%

\end{document}